\documentclass[sigconf, screen]{acmart}

\AtBeginDocument{%
  \providecommand\BibTeX{{%
    \normalfont B\kern-0.5em{\scshape i\kern-0.25em b}\kern-0.8em\TeX}}}

\copyrightyear{2022} 
\acmYear{2022} 
\setcopyright{acmlicensed}
\acmConference[ESEC/FSE '22]{Proceedings of the 30th ACM Joint European Software Engineering Conference and Symposium on the Foundations of Software Engineering}{November 14--18, 2022}{Singapore, Singapore}
\acmBooktitle{Proceedings of the 30th ACM Joint European Software Engineering Conference and Symposium on the Foundations of Software Engineering (ESEC/FSE '22), November 14--18, 2022, Singapore, Singapore}
\acmPrice{15.00}
\acmDOI{10.1145/3540250.3549123}
\acmISBN{978-1-4503-9413-0/22/11}

\usepackage{enumitem}
\usepackage{xspace}
\usepackage[normalem]{ulem}
\usepackage{subcaption}
\usepackage{makecell}
\usepackage{multirow}
\newcolumntype{C}[1]{>{\centering\arraybackslash}m{#1}}
\usepackage{color, colortbl}
\definecolor{Gray}{gray}{0.9}
\usepackage{balance}

\usepackage{listings}
\usepackage{microtype}
\newcommand{\todo}[1]{\textcolor{black}{#1}}
\newcommand{\tool}{\textsc{DeepPerf}\xspace}

\begin{document}

\title{Understanding Performance Problems in Deep Learning Systems}
\author{Junming Cao}
\authornote{Also with Shanghai Key Laboratory of Data Science, and Shanghai Collaborative Innovation Center of Intelligent Visual Computing.}
\affiliation{
\department{School of Computer Science}
\institution{Fudan University}
\city{Shanghai}
\country{China}
}

\author{Bihuan Chen}
\authornotemark[1]
\authornote{Bihuan Chen is the corresponding author.}
\affiliation{
\department{School of Computer Science}
\institution{Fudan University}
\city{Shanghai}
\country{China}
}

\author{Chao Sun}
\authornotemark[1]
\affiliation{
\department{School of Computer Science}
\institution{Fudan University}
\city{Shanghai}
\country{China}
}

\author{Longjie Hu}
\authornotemark[1]
\affiliation{
\department{School of Computer Science}
\institution{Fudan University}
\city{Shanghai}
\country{China}
}

\author{Shuaihong Wu}
\authornotemark[1]
\affiliation{
\department{School of Computer Science}
\institution{Fudan University}
\city{Shanghai}
\country{China}
}

\author{Xin Peng}
\authornotemark[1]
\affiliation{
\department{School of Computer Science}
\institution{Fudan University}
\city{Shanghai}
\country{China}
}

\begin{abstract}

Deep learning (DL) has been widely applied to many~domains.~Unique challenges in engineering DL systems are posed by the programming paradigm shift from traditional systems to DL systems,~and~performance is one of the challenges. Performance~problems~(PPs)~in~DL systems can cause severe consequences such as excessive resource~consumption and financial loss. While bugs in DL systems have been~extensively investigated, PPs in DL systems have hardly been explored. To bridge this gap, we present the first comprehensive~study~to~i)~characterize symptoms, root causes, and introducing and exposing~stages of PPs in DL systems developed~in~\textsc{TensorFLow}~and \textsc{Keras}, with~\todo{224} PPs collected from \todo{210} StackOverflow posts, and~to~ii)~assess~the~capability of existing performance analysis approaches in tackling~PPs, with a constructed benchmark of \todo{58}~PPs~in~DL~systems. Our~findings shed light on the implications on developing high-performance DL systems, and detecting and localizing PPs in DL systems.~To~demonstrate the usefulness of our findings, we develop a static checker~\tool to detect three types of PPs. It has detected \todo{488}~new~PPs~in~\todo{130} GitHub projects. \todo{105}~and~\todo{27}~PPs~have been confirmed and fixed. 

\end{abstract}



\maketitle


\section{Introduction}\label{sec:intro}

The advances in deep learning (DL) have attracted~an~increasing~interest in applying DL to various applications~in~both~industry~and academia, e.g., image processing, machine translation,~speech~recognition, medical diagnosis, self-driving cars, and robotics. DL~systems adopt a \textit{data-driven} programming paradigm,~where developers define a desired neural network that learns~the decision logic from~a large amount of training data. Differently, traditional~systems follow a \textit{logic-based} programming paradigm,~where developers directly~encode the decision logic in the source code. This paradigm shift poses unique challenges to engineering DL systems  \cite{amershi2019software, zhang2019empirical, han2020programmers, chen2020comprehensive}.

In particular, performance, as an important quality requirement,~is one of the challenges~in~engineering DL systems~\cite{zhang2019empirical}. It has a significant impact on the time and resources (e.g., GPU memory~and~power) required during the process pipeline (e.g., training~and~inference)~of DL systems~\cite{Menghani2021}. For example, the language model GPT-3~costs~millions of dollars for a single training run\footnote{https://lambdalabs.com/blog/demystifying-gpt-3/}. Performance~problems~(PPs) can slow down DL systems, consume excessive resources, hurt~user experience, cause financial loss, or threaten human lives. For example, many users suffered a significant slowdown of their~DL~systems after upgrading \textsc{TensorFlow} 1.x to \textsc{TensorFlow} 2.x, and hence~decided to switch to \textsc{PyTorch}\footnote{https://stackoverflow.com/questions/58441514/why-is-tensorflow-2-much-slower-than-tensorflow-1}. Moreover, performance questions of DL systems are recognized as the most difficult to answer~among~all questions of DL systems on StackOverflow~\cite{zhang2019empirical}.~Therefore,~it~is~necessary to study the characteristics of PPs in DL systems.

A lot of efforts have been recently made to extensively investigate the characteristics (e.g., symptoms, root causes, fixes and taxonomy) of general bugs~\cite{EmpiricalStudyTensorFlow2018, Islam2019, humbatova2019taxonomy, islam2020repairing} and specific bugs \cite{zhang2020empirical, chen2021empirical, zhang2021autotrainer, MLAPI2021}~in~DL~systems. However, these studies are not specifically designed for PPs, and thus only capture some partial characteristics~of PPs~in~DL~systems. In contrast,~PPs~have been widely studied~for traditional systems,~e.g.,~desktop or server applications~\cite{UnderstandingDetectingRealworld2012, zaman2012qualitative, nistor2013discovering, Song2014}, highly~configurable systems~\cite{EmpiricalStudyPerformance2016, he2020cp},~mobile applications~\cite{liu2014characterizing, linares2015developers}, database-backed web applications~\cite{yang2018not, yang2019view}, and JavaScript systems~\cite{selakovic2016performance}. However, PPs in DL systems could be different due to the programming paradigm shift from traditional systems to DL systems. In summary, the characteristics of PPs in DL systems are under-investigated.

To bridge this knowledge gap, we present the first comprehensive study to characterize PPs in DL systems developed in \textsc{TensorFlow} and \textsc{Keras} and to assess existing approaches in tackling~PPs.~To~this end, we first collect \todo{224} PPs from \todo{210} StackOverflow~posts,~and~manually investigate the PPs to characterize~their~symptoms (\textbf{RQ1}),~root causes (\textbf{RQ2}), and introducing and exposing~stages (\textbf{RQ3}). Based~on these~\todo{224} PPs, we manually build a benchmark~of~\todo{58} PPs~that cover most symptoms and root causes, and assess the capability~of~a~profiler in detecting PPs, the capability of a compiler in optimizing~PPs, and the capability of documentation in hinting PPs (\textbf{RQ4}).

\begin{itemize}[leftmargin=*]
\item \textbf{RQ1 Symptom:} what are the symptoms of PPs?

\item \textbf{RQ2 Root Cause:} what are the root causes of PPs?

\item \textbf{RQ3 Stage:} what are the stages of introducing and exposing~PPs?

\item \textbf{RQ4 Assessment:} how is the capability of existing performance analysis approaches in tackling PPs?
\end{itemize}
Through these research question analysis, we aim to provide useful findings for developers and researchers. For example, more~than~half of the PPs slow down DL systems,~and nearly one-third of the~PPs~consume either extremely low or high~resources. About half of~the~PPs are introduced by API~misuses, and root causes related to model,~data and hardware introduce more than one-third~of~the PPs. The most~bug-prone stages are data preparation,~environment setting, model building and training. The most bug-affecting stages are training~and~data preparation. \todo{40\%}~of~the PPs are not exposed~in~the~introducing stage. Existing approaches have a very limited capability in tackling PPs. Our findings provide implications for developers and researchers~on developing high-performance DL systems and detecting~and~localizing PPs in DL systems, e.g., performance-aware~techniques~to~recommend DL library APIs and~DL~models, static~techniques to model and estimate time cost~and~resource consumption~of DL systems, and rule-based techniques to detect and localize PPs in DL systems.

To demonstrate the usefulness of our findings, we develop~a~static checker, named \tool, that supports rule-based detection~of~three types of PPs derived from our study. We run \tool~against~\todo{1,108} GitHub projects with~more~than \todo{100} stars. \tool has detected~\todo{488} new PPs in \todo{130} of these projects with \todo{15} false positives. \todo{105} of these PPs have already been confirmed by the developers, and \todo{27} of them have already been fixed. Others are still waiting for confirmation.

In summary, this paper makes the following contributions.

\begin{itemize}[leftmargin=*]
\item We present the first comprehensive study to characterize \todo{224}~PPs in DL systems written in \textsc{TensorFlow} and \textsc{Keras},~and~to~assess existing approaches in tackling a contructed benchmark~of~\todo{58}~PPs.

\item We develop a~static checker, named \tool, to detect three~types of PPs, and detect \todo{488} new PPs in \todo{130} GitHub projects.
\end{itemize}

\section{Empirical Study Methodology}\label{sec:design}

We first introduce the design of our study, and then~present~our~data collection, data labeling, and benchmark construction process.

\subsection{Study Design}

Our goal is to understand PPs in DL systems. As DL systems~can~be built on top of various DL libraries, we limit our scope to DL systems developed in \textsc{TensorFlow} and \textsc{Keras}. We select \textsc{TensorFlow}~as~it is the most popular DL library on GitHub. 
We also include \textsc{Keras} because it is built on top of and tightly integrated with \textsc{TensorFlow} 2. We include \textsc{Keras} but do not distinguish~between \textsc{TensorFlow}~and \textsc{Keras} in our analysis because i) \textsc{Keras} is a frontend and should be used with a backend, and \textsc{TensorFlow} is the most popular backend, and ii) \textsc{TensorFlow}~and \textsc{Keras} are often tightly used together.


To achieve this goal, we propose the \todo{four} research questions~as~introduced in Sec.~\ref{sec:intro}. Our \textit{symptom analysis} in \textbf{RQ1} aims~to~understand the observable consequences of PPs. Our findings~from~\textbf{RQ1}~can~characterize the significance of PPs, and provide insights for developing PP detection approaches. Our~\textit{root cause analysis}~in~\textbf{RQ2}~aims~to~characterize the fundamental~reasons for the occurrence of PPs.~Our~findings from \textbf{RQ2} can provide~insights for designing PP localization~approaches. Our~\textit{stage analysis} in \textbf{RQ3} aims to locate~DL~pipeline~stages where~PPs~are~introduced and exposed, and measure the distance~between~exposing~stage~and introducing stage. Our findings from \textbf{RQ3} can~locate~the~bug-prone and bug-affecting stages that should~be~concerned, and reflect~the difficulty of PP localization. Our \textit{approach~assessment} in \textbf{RQ4} aims to quantitatively evaluate existing approaches in tackling PPs. Our findings from \textbf{RQ4}~can~reveal the necessity~of PP detection and localization approaches. Besides, our~findings can also provide hints to develop high-performance DL systems.


\subsection{Data Collection}


We collected PPs from a well-known~Q\&A site StackOverflow,~where world-wide developers can discuss software development~problems. Our PP collection process consists of the following three steps.

\textbf{Step 1: DL Post Selection.} We first selected posts related to~DL libraries \textsc{TensorFlow} and \textsc{Keras} by checking whether the tags~of~a post contain the keywords ``tensorflow'' and ``keras''. We also filtered posts that were created before 2018-01-01 to avoid usage~discussions about old versions of DL libraries that are usually~no~longer~used. At the time of selection (i.e., 2021-03-01), we obtained 61,169 DL posts. Then, we excluded posts that did not contain any source~code~in~question descriptions for the ease of our manual analysis. To focus~on~high-quality posts, we also excluded~posts that did not have an accepted answer or any answer whose votes were greater than two because questioners often commented that the problems had~been~solved,~but forgot to accept the answer. After this step, we had 18,730 DL posts.

\textbf{Step 2: PP Post Selection.} Instead of directly using performance-related keywords from the existing studies on PPs in traditional~systems (e.g., \cite{UnderstandingDetectingRealworld2012, zaman2012qualitative, nistor2013discovering, Song2014}), we derived a keyword set in the following way to achieve a wide and comprehensive coverage~of~PP~posts.~We first randomly sampled 100 posts with a tag of ``performance'' from 18,730 posts in Step 1. Then, we manually analyzed these posts~to~extract performance-related keywords, and added them to the set~of~keywords from existing studies. We continued this procedure of random sampling and manual analysis for another two rounds~until~no~new keyword was found; i.e., we sampled 300 posts, which achieved 95\% confidence level and 5.6\% confidence interval. Finally, we used~the derived keyword set to search question descriptions of the 18,730 posts in Step 1, which resulted in 742 candidate PP posts. We provide the full set of derived keywords at our replication site.


\textbf{Step 3: PP Identification.} We manually verified the 742 candidate PP posts to reduce noise that was not about PPs~in~DL~systems. For example, some posts might happen to have performance-related keywords, but did not discuss PPs; some posts actually discussed~the accuracy of DL models (because accuracy is often interchangeable with performance in the DL community, and we~align~with~the~SE community where performance is usually referred to as efficiency);~and some posts indeed discussed performance, but did not have a correct answer, which could not be used to understand the characteristics~of PPs. In particular, two of the authors separately inspected~each~candidate PP post to identify PPs. We used Cohen's~Kappa coefficient~to measure the agreement, and it reached 0.813.~A third author was~involved to resolve disagreements. Finally, we identified \todo{224} PPs from \todo{210} PP posts, of which~\todo{14} PP posts contained~two~PPs.~This~scale~is comparable to previous studies on PPs, e.g., 109 PPs in desktop~or~server applications \cite{UnderstandingDetectingRealworld2012} and 70 PPs in mobile applications~\cite{liu2014characterizing}.

 






\subsection{Data Labeling}\label{sec:label}

To answer \textbf{RQ1}, \textbf{RQ2} and \textbf{RQ3}, two of the authors labeled each~of~the \todo{224}~PPs with respect to \todo{three} dimensions: symptom, root cause,~and introducing and exposing stages. In particular, they started~with~the classification schema, used for labeling, from the existing general~DL bug~studies \cite{EmpiricalStudyTensorFlow2018, Islam2019, humbatova2019taxonomy, islam2020repairing}~and adapted~it by appending new~ones~and~excluding non-applicable ones. They separately read all post~contents, including the~title, question description, comments, answers,~and~reference~links~mentioned during discussion, to carefully label PPs.

Specifically, the symptom of a PP was determined if the questioner explicitly reported the symptom in the post. Otherwise,~it~was conservatively labeled as “\textit{Unknown}”. The root cause of a PP was~inferred from the buggy code version in the question and the fixed~code version (always existed) in the valid answer. The introducing stage~of a PP was determined by analyzing where its root cause was located, while the exposing stage of a PP was decided by analyzing~where its symptom~was exhibited. The introducing/exposing~stage~of~a~PP~was labeled as “\textit{Unknown}” if there was no clear indication in the post.~We provide actionable code of the final taxonomies for symptoms, root causes~and stages at our replication site.

The Cohen's~Kappa coefficient was 0.906, 0.772, 0.847 and 0.928~for the labeling of symptom, root cause, introducing stage and exposing stage. A third author~was~involved to resolve disagreements.~It~is worth mentioning that the manual effort, involved in our data collection and labeling procedure, required \todo{six} person-months.

\subsection{Benchmark Construction}\label{sec:benchmark}

To answer \textbf{RQ4}, we constructed a benchmark by reproducing~PPs. We reproduced PPs on a machine with a 16-core~Intel i7-7820X~CPU (3.60GHz), NVIDIA TITAN Xp GPU, 128GB~RAM and 1TB SSD.~Different PPs require different \textsc{TensorFlow} versions which further~require different CUDA Toolkit versions to support GPU. It is tricky~to install~different CUDA versions in the same physical machine.~Thus, we used \textsc{TensorFlow} Docker images. Only~NVIDIA~GPU~Driver was installed in the physical machine, and each docker container had its own CUDA Toolkit version. Finally, \textsc{TensorFlow} Docker~images ranging from version 1.12 to 1.15 and version~2.0~to~2.5~with~GPU support were covered to build our PP benchmark. 


We decided to sample some PPs from the \todo{224} PPs instead~of~trying to reproduce all \todo{224} PPs~due~to~the~large~effort~in~reproducing PPs from StackOverflow posts.~To~have a good coverage~of~symptoms and root~causes, we sampled~50\%~PPs~from each~set of~PPs~that~were caused by each~inner~category of root causes (see Sec.~\ref{sec:root}) while exhibiting~each~high-level category of symptoms (see Sec.~\ref{sec:symptom}).~For~each sampled PP, we reproduced it with the following three steps.

\textbf{Step 1: Decide \textsc{TensorFlow} Version.} If the \textsc{TensorFlow} version was shown in the post, we used it. If not,~we~checked~whether APIs specific to \textsc{TensorFlow} 1.x (e.g., \texttt{tf.Session}) or \textsc{TensorFlow} 2.x (e.g., \texttt{@tf.function}) existed in the post. If yes, we used the latest \textsc{TensorFlow} version~of 1.x (i.e., 1.15) or 2.x (i.e.,~2.5). If not, we used \textsc{TensorFlow} 2.5.

\textbf{Step 2: Complete Code Snippets.} As developers tend~to~only~include code fragments that are directly related to questions,~code~snippets in the post are often incomplete. Specifically, if the buggy~(or fixed) version~was executable, we completed the fixed (or buggy)~version based on it. Otherwise, we wrote missing code fragments~for buggy and fixed versions based on question description and answer. 

\textbf{Step 3: Reproduce Symptoms.} We executed the buggy~and~fixed version to reproduce symptoms reported in the post. We~may~change input data size, model parameters, etc. to reproduce~described~symptoms as our hardware environment might be different from the post. For PPs with out of memory errors, we set the maximum GPU memory limit with \texttt{tf.GPUOptions} such that the out of memory errors could be reproduced even on GPUs with a larger memory.

We successfully reproduced \todo{58} PPs from \todo{112} sampled PPs with four person-months effort.~The~main reasons~for~failed reproduction~are:~i) developers provide~very~incomplete code snippets in~the posts, making it difficult for us~to~complete the buggy~or fixed version, and ii) some PPs require specific hardware environments that are different from our machine. To foster future research on PPs in DL systems, we recorded for each PP in our benchmark its ~environment configuration, input data, buggy version,~fixed~version, \todo{performance~change after fixing}, and reproduction steps. 


\section{Empirical Study Results}\label{sec:results}

We present the results of the four research questions.

\subsection{Symptom Analysis (RQ1)}\label{sec:symptom}

\begin{figure}[!t]
\centering
\includegraphics[scale=0.40]{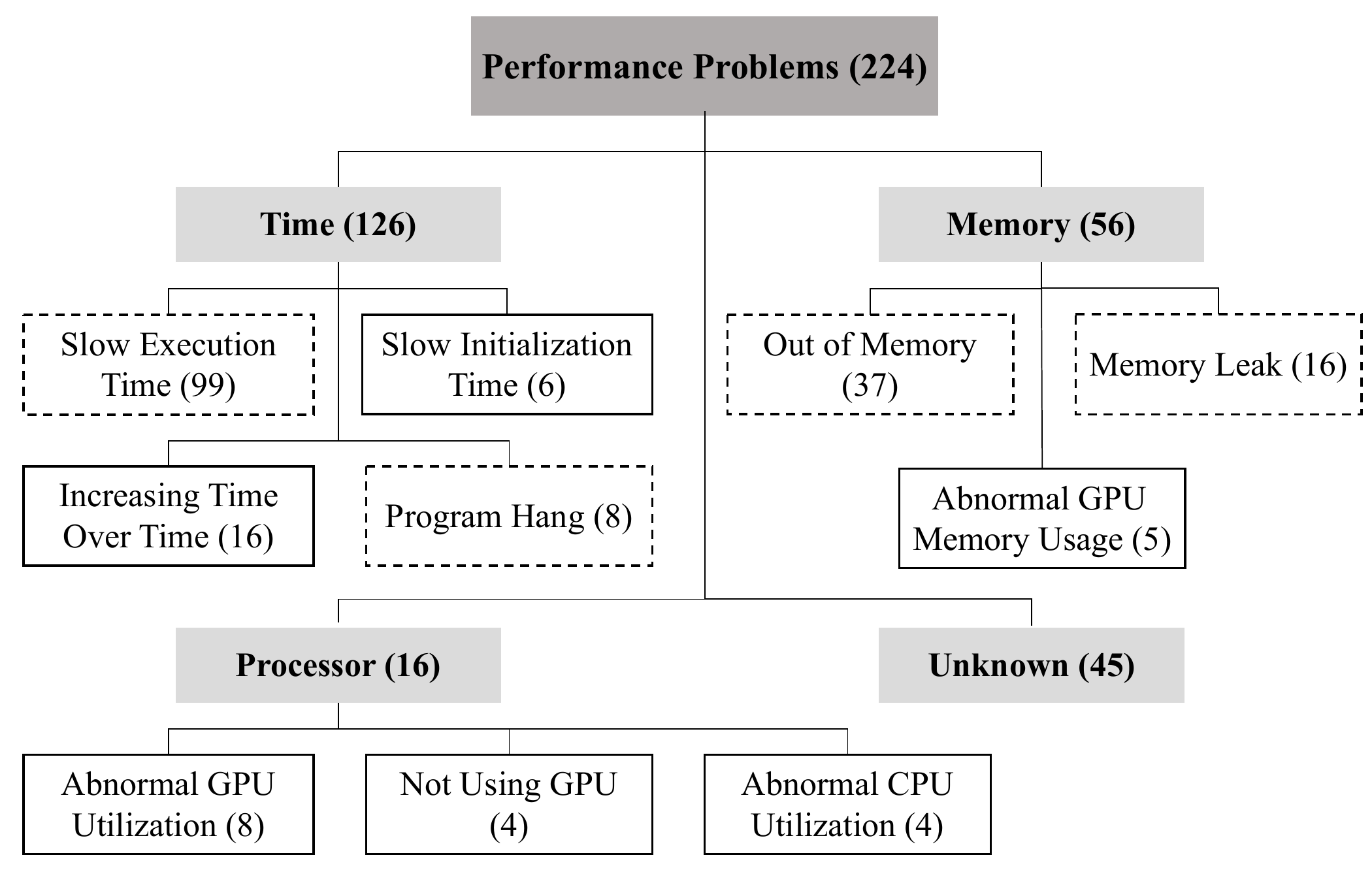}
\vspace{-20pt}
\caption{Taxonomy of PP Symptoms}\label{fig:symptoms}
\end{figure}

\begin{figure*}[!t]
\centering
\includegraphics[scale=0.51]{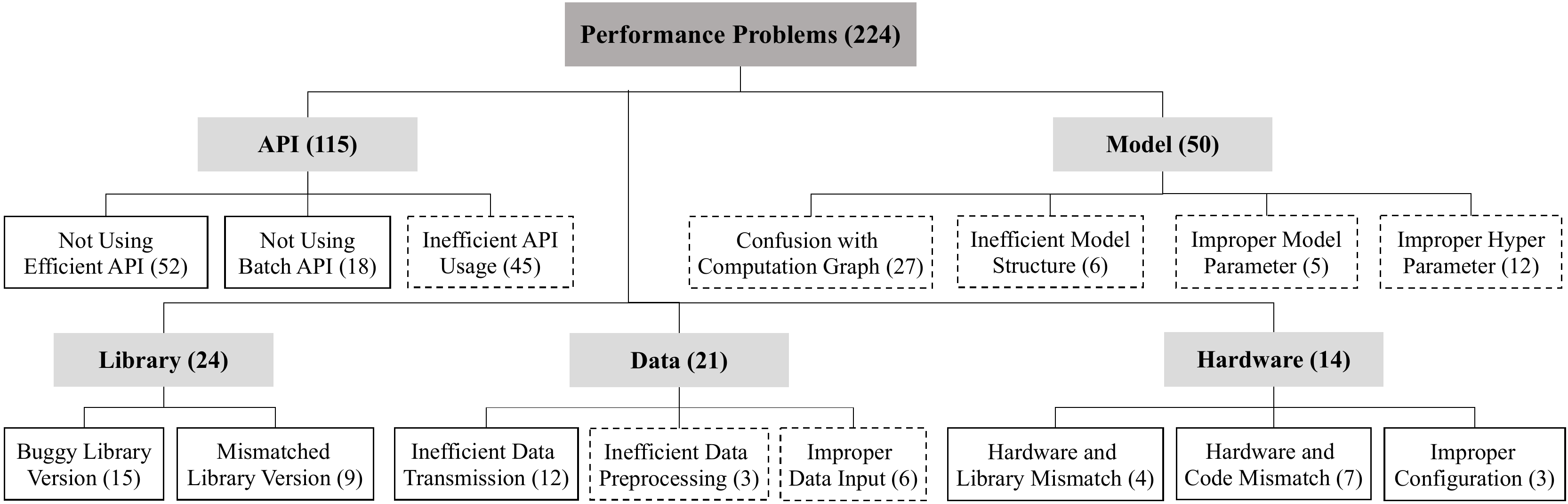}
\vspace{-5pt}
\caption{Root Causes of PPs in DL Systems}\label{fig:causes}
\end{figure*}

The taxonomy of PP symptoms is shown in Fig.~\ref{fig:symptoms}. It is organized~into three high-level categories (i.e., \textit{Time}, \textit{Memory} and \textit{Processor}) and~10 inner categories, which are exhibited by \todo{179~(79.9\%)}~of~the \todo{224}~PPs. The remaining \todo{45 (20.1\%)} PPs belong to the \textit{Unknown} category (defined in Sec. \ref{sec:label}). Notice that one PP can exhibit multiple symptoms.

\textbf{Time.} This category covers PPs exhibiting high time cost,~which accounts for the largest portion of PPs, i.e., \todo{126 (56.3\%)}. In particular, \todo{99 (44.2\%)} of the PPs manifest \textit{Slow Execution Time} during~the~execution of DL systems, including data~preparation, model building, training, evaluation, hyper parameter tuning, or prediction.~Further, \todo{16~(7.1\%)} of the PPs~exhibit~\textit{Increasing Time Over Time}; e.g.,~the~prediction time became longer and longer as the model ran\footnote{https://stackoverflow.com/questions/60267911/}. Moreover, \todo{6 (2.7\%)}~of~the PPs manifest \textit{Slow~Initialization Time} when DL systems are initialized before execution;~e.g.,~it~spent more than~80~seconds to import \textsc{TensorFlow}\footnote{https://stackoverflow.com/questions/49053434/}. DL systems~can~still work but slowly when exhibiting the above symptoms.~Differently, \todo{8 (3.6\%)} of the~ PPs result in \textit{Program Hang} that makes DL systems~cease~to respond to inputs, which is the most severe symptom.

\textbf{Memory.} This category includes PPs consuming RAM/GPU~memory abnormally, accounting for \todo{56 (25.0\%)} of the PPs. Specifically,~\textit{Out of Memory} is the most common as well as the most severe symptom, covering \todo{37 (16.5\%)} of the PPs. \textit{Memory Leak}, manifested~in~\todo{16~(7.1\%)} of the PPs, occurs when~the~memory usage keeps increasing,~and~may finally lead to out of memory errors. Moreover, \textit{Abnormal GPU~Memory Usage}, i.e., either unexpectedly high or low GPU memory usage, is exhibited in \todo{5 (2.2\%)} of the PPs. 

\textbf{Processor.} This category consists of PPs with abnormal CPU/GPU utilization, which accounts for \todo{16 (7.1\%)} of the PPs. In particular,~\textit{Abnormal GPU Utilization}, i.e., either unexpectedly high~or~low~GPU~utilization, is manifested in \todo{8 (3.6\%)} of the PPs. For example, the~GPU~utilization was only around 15\%, while the training time was slow~(each epoch took 40 to 50 seconds)\footnote{https://stackoverflow.com/questions/56795642/}. Moreover, DL systems may \textit{Not Use GPU}, leading to no speedup than when running on CPU, which~occurs in \todo{4 (1.8\%)} of the PPs. In addition, \textit{Abnormal CPU Utilization} is also exhibited~in \todo{4 (1.8\%)} of the PPs. 
 

\uline{\textbf{Summary.}} More than half of the PPs slow down DL systems,~and nearly one-third of the PPs consume either extremely low or high~resources like memory and processor. Such severe consequences~of~PPs motivate the significance of PPs. Moreover, only~four of the ten~symptoms, as highlighted in dotted rectangles in Fig.~\ref{fig:symptoms},~are~shared~with~the existing symptom taxonomies for general DL bugs~\cite{EmpiricalStudyTensorFlow2018, Islam2019}.~In other words, symptoms of PPs are quite different from those~of~general~DL bugs, and the existing studies on general DL bugs~only~capture~a~partial set of PPs, and thus PPs deserve a comprehensive investigation.

\subsection{Root Cause Analysis (RQ2)}\label{sec:root}

The taxonomy of PP root causes is reported in Fig.~\ref{fig:causes}. It is grouped~into five high-level categories (i.e., \textit{API}, \textit{Model}, \textit{Library},~\textit{Data} and \textit{Environment}) and 15 inner categories.

\textbf{API.} This category covers PPs caused by library API misuses.~This is the most common category and accounts for \todo{115 (51.3\%)}~of~the~PPs. Specifically, \textsc{TensorFlow}~and~\textsc{Keras} provide efficient APIs~for~achieving high performance, e.g., the \texttt{tf.data}~API for building efficient~input pipelines, and various operation~APIs~for efficient computation. However, developers often write their own~implementation~which~is often less efficient, but do \textit{Not Use} the corresponding \textit{Efficient API}~directly, potentially due to the unfamiliarity with APIs.~This causes~\todo{52 (23.2\%)} of the PPs. For example,~a developer wrote a \texttt{for}~loop~to~perform concatenation on a set of images, which could be efficiently achieved by the \texttt{map} API from \texttt{tf.data.Dataset}\footnote{https://stackoverflow.com/questions/63002205/}. Moreover, \textsc{TensorFlow}~and~\textsc{Keras} provide various batch processing APIs for high performance, e.g., data loading, training, evaluation or prediction~in a batch mode. However, developers might~\textit{Not~Use~a~Batch~API},~and some even implement batch processing by themselves, which causes \todo{18 (8.0\%)} of the PPs. For example, a developer loaded~a large~data~set into GPU memory all at once, causing an out of memory error\footnote{https://stackoverflow.com/questions/59456128/}.~The \texttt{flow\_from\_directory} API in \textsc{Keras} can solve~this PP by dynamically loading a batch of data from the specified directory. Notice that \textit{Not Using Batch API} is a sub-category of \textit{Not Using Efficient API}, and we treat it separately due to its high frequency.~In~the~previous two root causes, developers are mostly unaware of the efficient or batch APIs. However, even when developers are aware of some APIs, they might not fully understand their performance~characteristics, and write \textit{Inefficient API Usage}, which causes \todo{45 (20.1\%)} of the PPs. Fig.~\ref{fig:example-1} shows an example of inefficient API usage, where~a developer called the \texttt{map} API before the \texttt{batch} API, and did~not~pass the \texttt{num\_parallel\_calls} argument to \texttt{map}\footnote{https://stackoverflow.com/questions/53424152/}, leading to a long training time. To speed up, \texttt{map} should be called after \texttt{batch} to reduce the number of times the mapped function \texttt{\_batch\_parser} is called, and \texttt{num\_parallel\_calls} should be passed to enable parallelism.

\begin{figure}[!t]
\flushleft 
\begin{lstlisting}
- def _parser(record):][3]
+ def _batch_parser(record_batch):
-   parsed = tf.parse_single_example(record, _keys_to_map)
+   parsed = tf.parse_example(record_batch, _keys_to_map)
    return parsed['d'], parsed['s']

  def init_tfrecord_dataset():
    files_train = glob.glob(DIR_TFRECORDS + '*.tfrecord')
    random.shuffle(files_train)

    with tf.name_scope('tfr_iterator'):
      # define data from randomly ordered files
      ds = tf.data.TFRecordDataset(files_train)      
      # select elements randomly from the buffer
      ds = ds.shuffle(buffer_size=10000)
-     # map them based on tfrecord format
-     ds = ds.map(_parser)
      # group elements in batch
      ds = ds.batch(BATCH_SIZE, drop_remainder=True) 
+     # map batches based on tfrecord format
+     ds = ds.map(_batch_parser, num_parallel_calls=4)             
      # iterate infinitely         
      ds = ds.repeat()                               

      # initialize the iterator
      return ds.make_initializable_iterator()        
\end{lstlisting}
\vspace{-10pt}
\caption{Inefficient API Usage Before and After Fix}\label{fig:example-1} 
\end{figure}

\textbf{Model.} This category consists of PPs that are related to DL~models, which is the second most common category, accounting~for~\todo{50 (22.3\%)} of the PPs. In particular, developers may have \textit{Confusion with Computation Graph} because of the unfamiliarity with~the~programming model in \textsc{TensorFlow} and \textsc{Keras}, which~causes~\todo{27~(12.1\%)}~of~the PPs. A typical confusion is with the programming model of \textsc{TensorFlow} 1.x, which is to first build a dataflow computation graph~and~then run it repeatedly with inputs being fed to and outputs being fetched from the graph. Developers often mix the graph construction into~the graph execution. As a result, nodes are repeatedly added to the~graph, and the graph execution becomes slower and slower. An example\footnote{https://stackoverflow.com/questions/53137115/} is shown in Fig.~\ref{fig:example-2}, where Line 14--16 builds the graph and should~be moved out of the execution loop to Line  6--8. Another~common confusion is with the usage of session, which owns resources~like~queues and variables. However, developers repeatedly create a session~in~the graph execution loop without reusing, or forget to close the session. The example in Fig.~\ref{fig:example-2} also forgets to close the session, and the fix~is to use the session as a context manger at Line 11 that~will~automatically close the session. A typical confusion in \textsc{TensorFlow}~2.x~is~with the \texttt{@tf.function} decorator, which accelerates the decorated function by running it in graph mode instead of in eager mode.~However, developers often do not know where to add the decorator~and~how~to design the decorated function to get real speedup. Further, developers design an \textit{Inefficient Model Structure} (e.g., missing convolution and pooling layers before the flatten layer to have~too~many~weights) or set \textit{Improper Model Parameter} (e.g., a large kernel size in a convolution layer to cause a long training time). These two categories~respectively cause \todo{6 (2.7\%)} and \todo{5 (2.2\%)} of the PPs.~Moreover,~developers also set \textit{Improper Hyper Parameter}, e.g., a~large~batch~size~to~cause an~out of memory error or a small batch size to cause a long training time. This category causes \todo{12 (5.4\%)} of the PPs.

\begin{figure}[!t]
\flushleft 
\begin{lstlisting}
  inp = tf.constant([[1.,1.]])
  out = tf.constant([[1.,0.]])
  weight = tf.Variable([[1.,1.], [1.,1.]])

  optimizer = tf.train.GradientDescentOptimizer(0.1)
+ y = tf.matmul(inp, weight)
+ loss = (out[0][0] - y[0][0])*2 + (out[0][1] - y[0][1])*2
+ train = optimizer.minimize(loss)

- sess = tf.Session()
+ with tf.Session() as sess:
    sess.run(tf.global_variables_initializer())
    for epoch in range(1000):
-     y = tf.matmul(inp, weight)
-     loss = (out[0][0] - y[0][0])*2 + (out[0][1] - y[0][1])*2
-     sess.run(optimizer.minimize(loss))
+     sess.run(train)
\end{lstlisting}
\vspace{-10pt}
\caption{Graph Confusion Before and After Fix}\label{fig:example-2} 
\end{figure}

\textbf{Library.} This category refers to PPs caused by problems~of~DL~libraries, accounting for \todo{24 (10.7\%)} of the PPs. Specifically, \todo{15 (6.7\%)}~of the PPs are caused by \textit{Buggy Library Version}; i.e., DL systems~themselves are correctly written, but trigger the PPs in DL libraries.~For example, repeated calls to \texttt{model.predict} (e.g., in a loop) resulted~in a memory leak\footnote{https://stackoverflow.com/questions/60267911/}, due to a memory leak~persisting across multiple versions of \textsc{TensorFlow}\footnote{https://github.com/tensorflow/tensorflow/issues/34579/}. These \todo{15} PPs trigger the PPs~in~\todo{12}~distinctive APIs. It is~non-trivial to detect such PPs as we do not~have~a full~list~of APIs with PPs in each DL library version. Moreover,~\textit{Mismatched Library~Version}~causes~\todo{9~(4.0\%)} of the PPs, as version restrictions have to be satisfied~for~full~GPU~usage. For example,~\textsc{TensorFlow} 1.x is not~fully~supported on \textsc{CUDA} 11.1, resulting in a long time to start the training\footnote{https://stackoverflow.com/questions/64462347/}.


\textbf{Data.} This category covers PPs related to data processing,~accounting for \todo{21 (9.4\%)} of the PPs. Specifically, developers may~write \textit{Inefficient Data Transmission}, e.g., loading input data over the network during training but not directly copying them to the local storage,~or storing weight data in CPU which causes the weights copied~to~GPU and the gradients copied back to CPU in each training iteration.~This category accounts for \todo{12 (5.4\%)} of the PPs. Further, developers may implement \textit{Inefficient Data Preprocessing} (e.g., lack of image normalization before changing an image to a tensor), which causes~\todo{3~(1.3\%)} of the PPs. Moreover, \textit{Improper Input~Data} (e.g., improper data~format or size that consumes excessive resources) causes \todo{6 (2.7\%)}~of~the PPs. For example, images with unnecessarily high resolution were loaded, causing an out of memory error\footnote{https://stackoverflow.com/questions/50742757/}.

\begin{figure*}[!t]
\centering
\begin{subfigure}[b]{0.48\textwidth}
\centering
\includegraphics[width=0.99\textwidth]{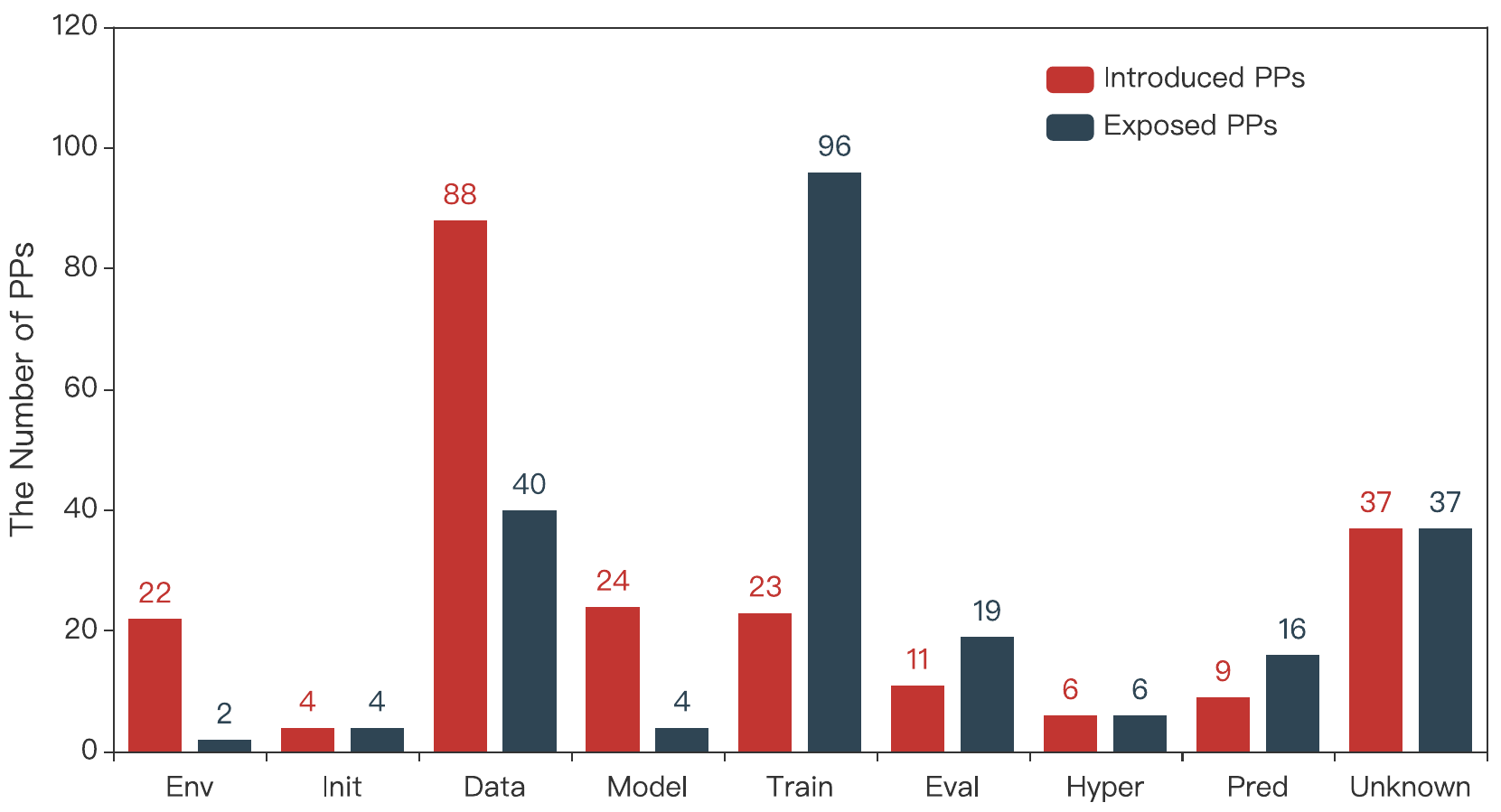}
\vspace{-2pt}
\caption{Introduced and Exposed PPs in Each Stage}
\label{fig:stage-bug}
\end{subfigure}
\begin{subfigure}[b]{0.485\textwidth}
\centering
\includegraphics[width=0.99\textwidth]{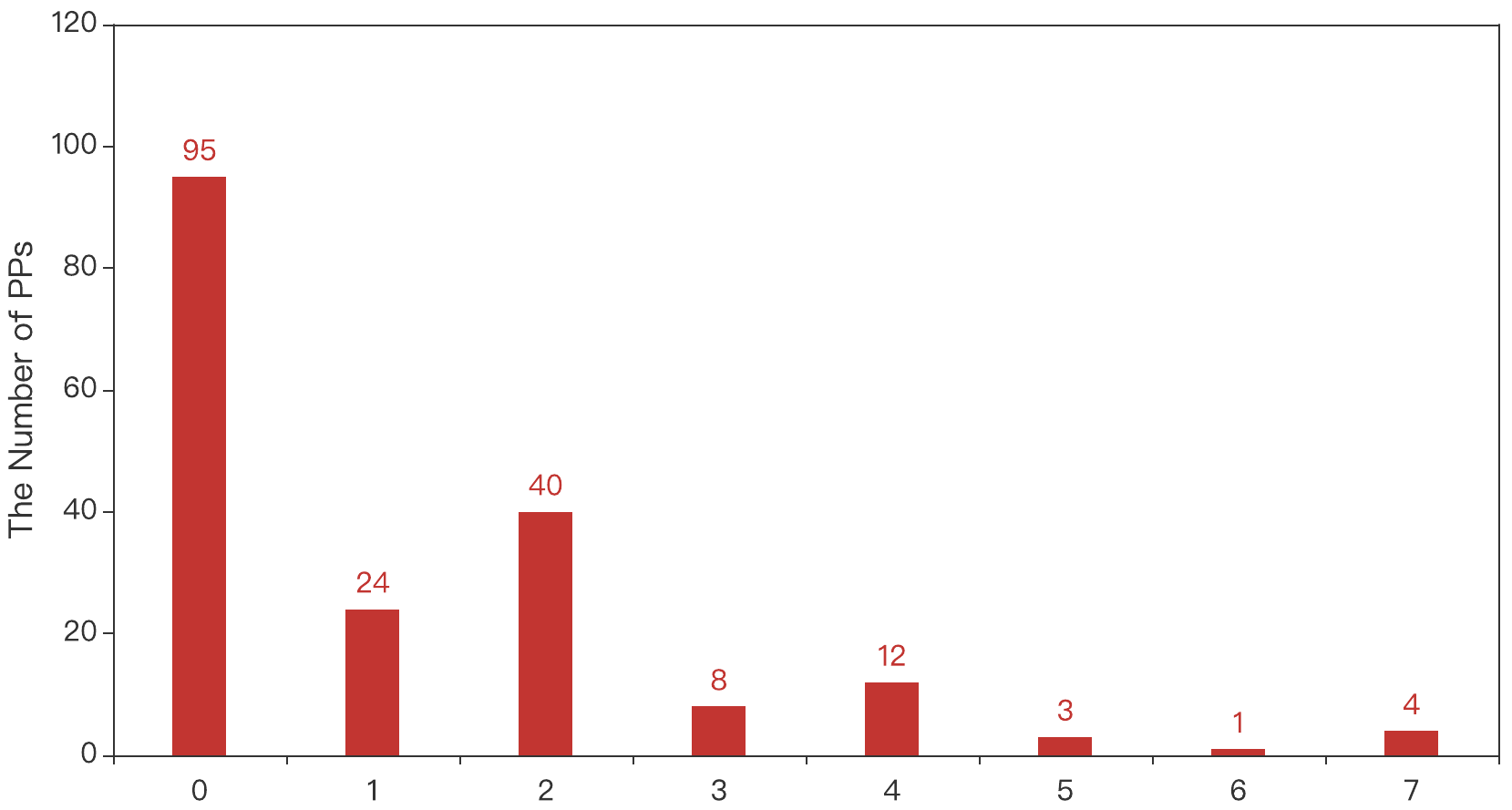}
\vspace{-2pt}
\caption{Distance between Exposing Stage and Introducing Stage}
\label{fig:stage-distance}
\end{subfigure}
\vspace{-10pt}
\caption{The Exposing Stage and Introducing Stage of PPs and their Distance}
\end{figure*}

\begin{figure*}[!t]
\centering
\begin{subfigure}[b]{0.48\textwidth}
\centering
\includegraphics[width=0.99\textwidth]{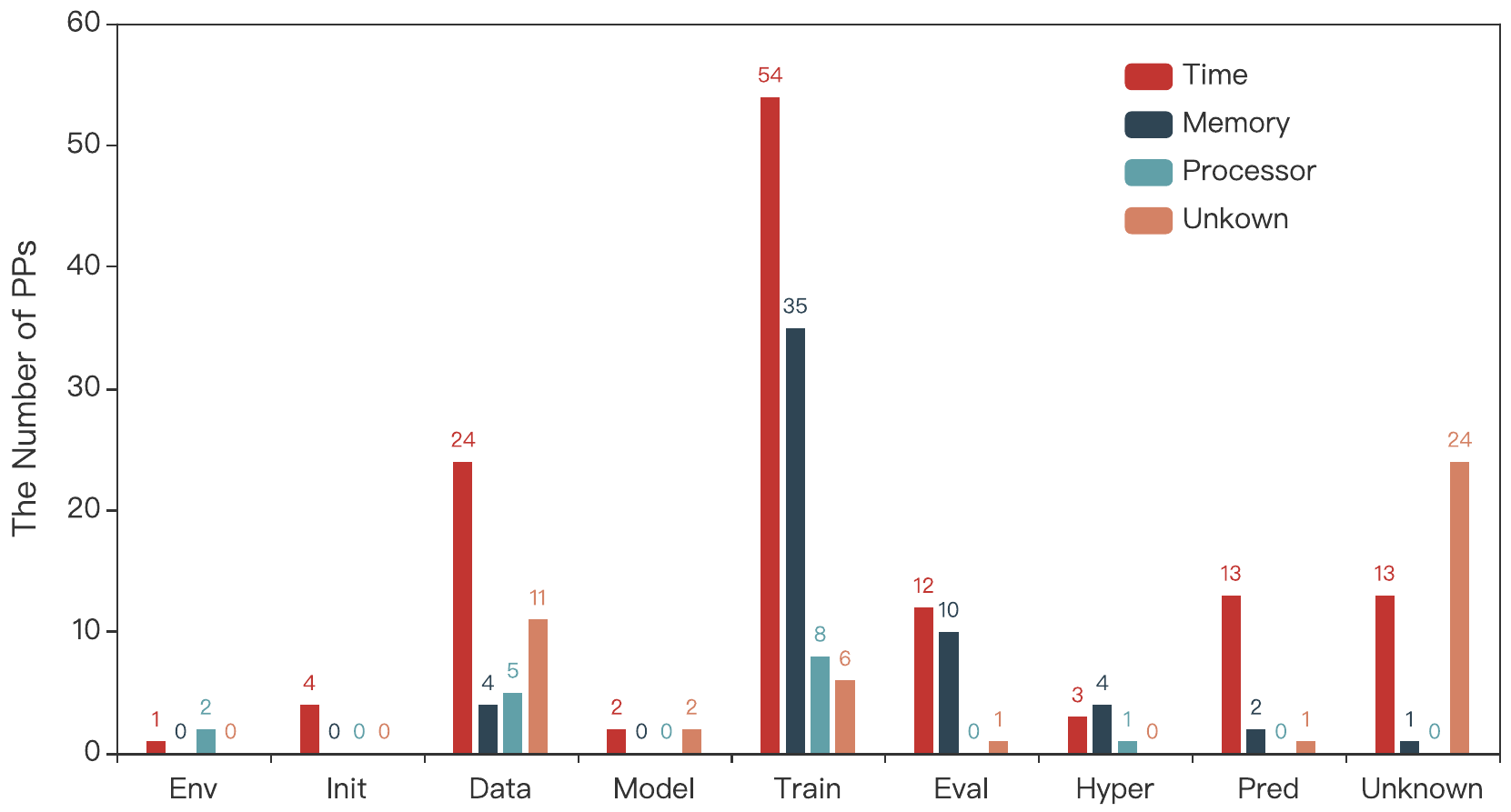}
\vspace{-2pt}
\caption{Symptoms of the PPs Exposed in Each Stage}
\label{fig:stage-symptom}
\end{subfigure}
\begin{subfigure}[b]{0.48\textwidth}
\centering
\includegraphics[width=0.99\textwidth]{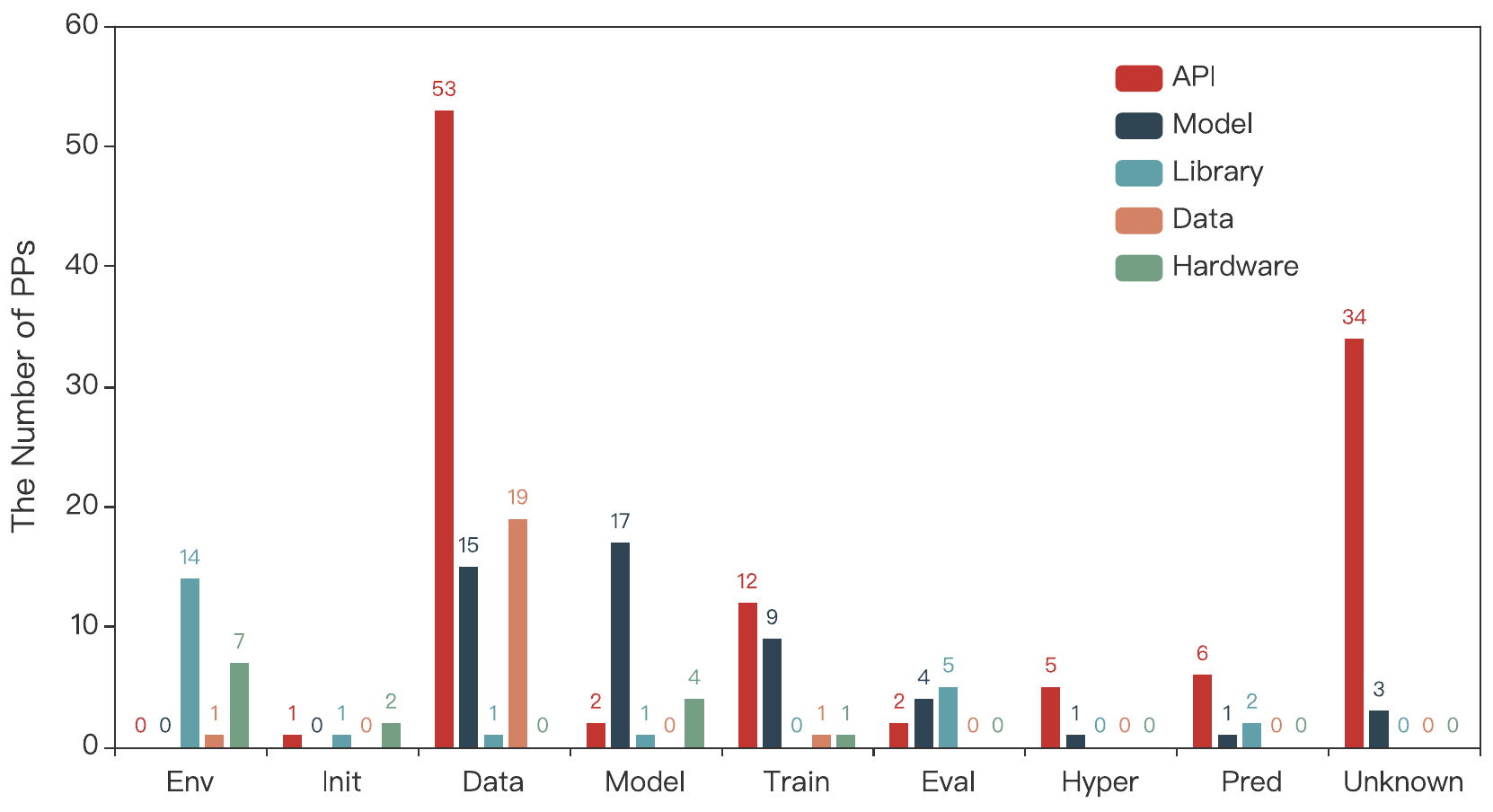}
\vspace{-2pt}
\caption{Root Causes of the PPs Introduced in Each Stage}
\label{fig:stage-cause}
\end{subfigure}
\vspace{-10pt}
\caption{Correlation between Symptoms and Exposing Stages and between Root causes and Introducing Stages}
\end{figure*}

\textbf{Hardware.} This category covers PPs related to hardware issues, accounting for \todo{14 (6.3\%)} of the PPs. Specifically, hardware may only support part of the DL library versions, and hence \textit{Hardware~and~Library Mismatch} causes \todo{4 (1.8\%)} of the PPs. For example, a GPU~with compute capability 6.1 is not supported in \textsc{TensorFlow} 2.3 which requires a GPU with compute capability 7.0\footnote{https://stackoverflow.com/questions/63602858/}. Further,~to~utilize~the full acceleration capability of TPU, DL systems often need specific code design. Thus, \textit{Hardware and Code Mismatch} causes \todo{7 (3.1\%)} of the PPs. For example, to use Colab TPU, a DL model need~to~be~explicitly converted to a TPU compatible version; if not, the training becomes extremely slow\footnote{https://stackoverflow.com/questions/58670563/}. Moreover, hardware~need proper configuration to achieve full utilization, especially for distributed training. Thus, \textit{Improper Configuration} causes \todo{3 (1.3\%)} of the PPs. For example, the \texttt{tf.distribute.Strategy} API should be used to properly configure and allocate multiple GPUs\footnote{https://stackoverflow.com/questions/59074659/}.

\uline{\textbf{Summary.}} About half of the PPs are introduced by API~misuses. Model, data and hardware, i.e., the enabling characteristics~of~DL systems, introduce more than one-third of the PPs. DL libraries~also~introduce one-tenth of the PPs. These diverse sources of root causes~increase the complexity of PP localization. Moreover, only seven~of~the 15 root causes, as shown in dotted rectangles in Fig.~\ref{fig:causes},~are~the~same~to the previous root cause taxonomies for general DL bugs~\cite{EmpiricalStudyTensorFlow2018, Islam2019, humbatova2019taxonomy}. These differences owe to the fact that our study is focused~on~the~performance of DL systems, while the previous studies are mainly~concentrated on the functionality of DL systems.

\subsection{Stage Analysis (RQ3)}

Islam et al.~\cite{Islam2019} classify the pipeline of DL systems into six stages,~i.e., \textit{Data Preparation}, \textit{Model Building},~\textit{Training}, \textit{Evaluation}, \textit{Hyper Para-meter Tuning} and \textit{Prediction}, in their study on general DL bugs.~We~consider them as the execution stages of DL systems, and add two new stages, found in our data labeling, before them.~The first~newly~added stage~is~\textit{Environment Setting}, where~DL~environment like libraries and hardware are installed~and~configured. The second one is \textit{Initialization}, where the DL system~is~initialized (e.g., importing libraries and initializing parameters) before starting the execution stages. 

Fig.~\ref{fig:stage-bug} reports the number of PPs introduced and exposed~in~each stage, where the stage name on the $x$-axis is simplified to the initial letters. Data preparation is the most bug-prone stage, which~is~blamed in \todo{88 (39.3\%)} of the PPs. Environment setting, model building~and~training are the second most bug-prone stages, respectively~causing~about \todo{10\%}~of the PPs. Hence, developers should~pay~more~attention~to~these stages to avoid the introduction of PPs, while~automated PP localization approaches should be specifically developed~for~these~stages. The other stages are less bug-prone, respectively introducing~at~most \todo{5\%} of the PPs. On the other hand, training~and data preparation~are the two most~bug-affecting stages, where \todo{96 (42.9\%)} and \todo{40~(17.9\%)}~of the PPs are respectively exposed. Thus, developers should focus more~efforts on these two stages to optimize their performance, while~automated PP detection approaches should be specifically developed~for these two stages. Around \todo{7\%} of the PPs are respectively exposed~until~the~evaluation and prediction stages. The other stages are less~bug-affecting, respectively exposing~at~most~\todo{3\%}~of~the~PPs. 

Further, data preparation introduces more PPs than exposed.~This difference is more severe in the other two earlier pipeline~stages, i.e., environment setting and model building. About \todo{62\%} of the~PPs~are introduced in the earlier four pipeline~stages, about \todo{61\%} of which~are exposed in the~later~four pipeline stages. The other way around,~training exposes more PPs than introduced. This difference holds~in~the other two later pipeline stages, i.e., evaluation and prediction.~Nearly \todo{61\%} of the PPs are exposed in the later four pipeline stages.~Thus,~PPs should be proactively detected and localized before severe consequences occur so as to reduce time cost and resource consumption. Besides, for each PP, we measure the distance between~its~exposing stage and introducing stage, which is used as an indicator~of~the~difficulty of PP localization. Intuitively, the~larger~the~distance,~the~more difficult to localize a PP from its symptom~to~root~cause.~As~shown~in Fig.~\ref{fig:stage-distance}, \todo{95 (42.4\%)} of the PPs are exposed~and~introduced~in~the~same stage, while \todo{92 (41.1\%)} of the PPs cannot be exposed in the introducing~stage. Specifically, \todo{68 (30.4\%)} of the PPs are exposed~two~stages later. Extremely, \todo{4 (1.8\%)} of the PPs are exposed seven stages later;~i.e., they are introduced in the first stage but exposed in the last stage. Hence, PP localization is challenging for a considerable amount~of~PPs. 

Moreover, we investigate the symptom distribution~of~the~PPs~exposed in each stage, which is shown in Fig.~\ref{fig:stage-symptom}. This distribution~helps pinpoint the potentially useful performance indicators for detecting PPs exposed in different stages. For example, time-related indicators can be valuable to~detect PPs exposed in initialization,~data~preparation and prediction, because the most common symptom of the~PPs exposed in these stages is under the category of \textit{Time}. Similarly,~we report the root~cause distribution of the PPs introduced in each~stage in Fig.~\ref{fig:stage-cause}. This distribution helps hint the potential technical solutions to localize PPs introduced in different stages.~For~example,~the most frequent root cause of the PPs introduced in most stages is under the category~of~\textit{API}, and hence API~misuse detection could~be developed to localize PPs introduced in these stages.

\uline{\textbf{Summary.}} The most bug-prone stages are data preparation,~envi-ronment setting, model building and training, which introduce~nearly \todo{70\%} of the PPs. The most bug-affecting stages are training~and~data preparation, which expose around \todo{60\%} of the PPs. Nearly \todo{40\%}~of~the PPs cannot be exposed in the introducing stage. Moreover, we introduce two new stages that are not covered in the previous~stage~analysis for general DL bugs~\cite{Islam2019}, and investigate the introducing~and~exposing stages that are not distinguished in the previous study~\cite{Islam2019}.



\begin{table*}
\centering
\small
\caption{Benchmark PPs across Root Causes and Symptoms and Assessment Results}\label{table:benchmark}
\vspace{-10pt}
\begin{tabular}{|c|*{4}{C{4.4em}|}c|*{5}{C{2.5em}|}}
\hline
\multirow{2}{*}{\textbf{Root Cause}} & \multicolumn{4}{c|}{\textbf{Symptom}} & \multirow{2}{*}{\textbf{Total}} & \multicolumn{2}{c|}{\textbf{Profiler}} & \multicolumn{2}{c|}{\textbf{XLA}} & \multicolumn{1}{c|}{\textbf{Doc.}} \\\cline{2-5}\cline{7-11}
& \textbf{Time} & \textbf{Memory} & \textbf{Processor} & \textbf{Unknown}  & & \textbf{App.} & \textbf{Par.} & \textbf{App.} & \textbf{Par.} & \textbf{App.} \\\hline\hline
\rowcolor{Gray}
\textbf{API} & \textbf{18 (54)} & \textbf{4 (23)} & \textbf{0 (10)} & \textbf{10 (38)} & \textbf{32 (115)} & \textbf{6} & \textbf{1} & \textbf{17} & \textbf{3} & \textbf{9} \\\hline
Not Using Efficient API & 8 (19) & 0 (1) & 0 (5) & 10 (31) & 17 (52) & 1 & 1 & 14 & 3 & 2\\
Not Using Batch API & 1 (6) & 2 (9) & 0 (1) & 0 (2) & 3 (18) & 2 & 0 & 0 & 0 & 0\\
Inefficient API Usage & 9 (29) & 2 (13) & 0 (4) & 0 (5) & 7 (45) & 3 & 0 & 3 & 0 & 7\\\hline\hline
\rowcolor{Gray}
\textbf{Model} & \textbf{10 (30)} & \textbf{7 (19)} & \textbf{0 (0)} & \textbf{2 (5)} & \textbf{17 (50)} & \textbf{8} & \textbf{1} & \textbf{7} & \textbf{1} & \textbf{2} \\\hline
Confusion with Computation Graph & 7 (22) & 2 (4) & 0 (0) & 1 (3) & 9 (27) & 2 & 0 & 6 & 1 & 0\\
Inefficient Model Structure & 0 (2) & 1 (2) & 0 (0) & 1 (2) & 2 (6) & 1 & 0 & 1 & 0 & 0 \\
Improper Model Parameter & 2 (2) & 3 (4) & 0 (0) & 0 (0) & 4 (5) & 4 & 1 & 0 & 0 & 0\\
Improper Hyper Parameter & 1 (4) & 1 (9) & 0 (0) & 0 (0) & 2 (12) & 1 & 0 & 0 & 0 & 2\\\hline\hline
\rowcolor{Gray}
\textbf{Library} & \textbf{4 (15)} & \textbf{4 (9)} & \textbf{0 (3)} & \textbf{0 (1)} & \textbf{6 (24)} & \textbf{0} & \textbf{0} & \textbf{0} & \textbf{0} & \textbf{0}\\\hline
Buggy Library Version & 4 (8) & 4 (8) & 0 (0) & 0 (1) & 6 (15) & 0 & 0 & 0 & 0 & 0\\
Mismatched Library Version & 0 (7) & 0 (1) & 0 (3) & 0 (0) & 0 (9) & -- & -- & -- & -- & --\\\hline\hline
\rowcolor{Gray}
\textbf{Data} & \textbf{2 (14)} & \textbf{0 (4)} & \textbf{1 (3)} & \textbf{1 (1)} & \textbf{3 (21)} & \textbf{1} & \textbf{0} & \textbf{0} & \textbf{0} & \textbf{1}\\\hline
Inefficient Data Transmission & 1 (10) & 0 (1) & 1 (2) & 0 (0) & 1 (12)& 0 & 0 & 0 & 0 & 1\\
Inefficient Data Preprocessing & 0 (1) & 0 (1) & 0 (0) & 1 (1) & 1 (3) & 1 & 0 & 0 & 0 & 0 \\
Improper Data Input & 1 (3) & 0 (2) & 0 (1) & 0 (0) & 1 (6) & 0 & 0 & 0 & 0 & 0  \\\hline\hline
\rowcolor{Gray}
\textbf{Hardware} & \textbf{0 (13)} & \textbf{0 (1)} & \textbf{0 (0)} & \textbf{0 (0)} & \textbf{0 (14)} & -- & -- & -- & -- & -- \\\hline\hline
\rowcolor{Gray}
\textbf{Total} & \textbf{34 (126)} & \textbf{15 (56)} & \textbf{1 (16)} & \textbf{13 (45)} & \textbf{58 (224)} & \textbf{15} & \textbf{2} & \textbf{24} & \textbf{4} & \textbf{12} \\\hline
\end{tabular}
\end{table*}

\subsection{Approach Assessment (\textbf{RQ4})}\label{sec:assessment}

To~the~best of our knowledge, there is no PP detection and localization approach for DL systems. Notice that performance analysis approaches in \cite{qi2016paleo, gao2020estimating} can estimate performance metrics (i.e., time and GPU memory), but cannot directly pinpoint PPs. Based~on~their estimation, either automated approaches need to be further designed or developer experience need to be relied on to identify PPs. Therefore, we do not use them. Thus, we select and assess~the~following~three~typical performance analysis approaches, which can~be~used by~developers~to~improve the performance of DL systems.

\begin{itemize}[leftmargin=*]
\item \textsc{TensorFlow} Profiler\footnote{https://tensorflow.org/guide/profiler}: It is built on top of NVIDIA CUDA Profiling Interface to track the performance of \textsc{TensorFlow} models.~It visualizes the time cost and resource consumption of various~\textsc{TensorFlow} operations in the model, finds performance~bottlenecks, and recommends~best~practices to improve performance. Differently, general python profiling tools (e.g., cProfile and memory\_profiler) can only measure performance metrics, but cannot directly pinpoint PPs. Therefore, we do not use them.


\item XLA (Accelerated Linear Algebra)\footnote{https://tensorflow.org/xla}: It is a domain-specific compiler that can accelerate \textsc{TensorFlow} models. Each \textsc{TensorFlow} operation is executed~by a precompiled GPU kernel implementation. XLA can compile the~\textsc{TensorFlow} graph into a sequence of computation kernels generated specifically for the given model, and fuse the kernels to avoid memory operations between~the~execution of different kernels to improve the performance~\cite{Li2021DLCompiler}.


\item \textsc{TensorFlow} Documentation: It includes~all \textsc{TensorFlow}~API~documentation\footnote{https://www.tensorflow.org/versions/r2.5/api\_docs} and performance guide\footnote{https://www.tensorflow.org/guide} where developers can~find hints about performance problems and optimization solutions. 

\end{itemize}

Generally, we assess each technique in two dimensions: i) whether a technique is applicable to a PP (or whether a PP is in the capability scope of a technique), and ii) whether a technique can~solve~a~PP.~The assessment results on our benchmark (see Sec.~\ref{sec:benchmark}) are shown~in~the last five columns in Table~\ref{table:benchmark}. The first six columns of Table~\ref{table:benchmark} show the number of reproduced~PPs across root causes~and~symptoms, where the number in parentheses is the total number of PPs. They~cover~all root~causes except for \textit{Mismatched Library Version} and the three~hardware relevant~root~causes. They cover all high-level symptoms,~but achieve a relatively~low coverage of processor relevant symptoms.

As shown in the seventh column of Table~\ref{table:benchmark}, \textsc{TensorFlow}~Profiler is only applicable to \todo{15 (25.9\%)} PPs, but is not applicable~to~the~others for two reasons. First,~\textsc{TensorFlow} Profiler requires a \textsc{TensorFlow} version of at least 1.14. However, some PPs are~reproduced with a lower version.~Second,~\textsc{TensorFlow}~Profiler requires~a~full training or evaluation process to track the performance, which is not always available for the PPs in our benchmark. Moreover,~of~these~\todo{15}~PPs, \textsc{TensorFlow}~Profiler fails to finish profiling because of out~of~memory errors for \todo{9} PPs, and does not raise any warning or raises~a~false warning for \todo{4} PPs. Hence, we consider these \todo{13} PPs as not solved~by \textsc{TensorFlow}~Profiler. For the remaining \todo{2} PPs, \textsc{TensorFlow}~Profiler either raises a warning but suggests a fix that achieves a smaller performance improvement than our fixed version in the benchmark, or helps detect the PP by reporting the most time-consuming operation but fails to raise a warning and suggest a fix. Thus,~we~consider these \todo{2} PPs as partially solved by \textsc{TensorFlow}~Profiler,~as~reported in the eighth column of Table~\ref{table:benchmark}. These results demonstrate~that~\textsc{TensorFlow} Profiler has limited capability in tackling PPs.


As presented in the ninth column of Table~\ref{table:benchmark}, XLA is applicable~to \todo{24 (41.4\%)} PPs. There are two reasons that XLA is not applicable to the others. First, XLA uses just-in-time (JIT)~compilation.~However, compilation errors might occur for some PPs in our benchmark.~Second, XLA is designed for optimizing the performance~of~\textsc{TensorFlow} models. Thus, it is not applicable to PPs whose root causes~are not related to \textsc{TensorFlow} operations or computation graphs. Furthermore, of these \todo{24} PPs, XLA only improves the performance~for \todo{4} PPs but still achieves a smaller performance improvement than our fixed version in the benchmark.~This~is reasonable because~XLA~is actually not aware of the PPs, but optimizes performance by fusing nodes in computation graphs, while our fixed version reduces the number of nodes in computation graphs. Hence, we consider these~\todo{4} PPs as partially solved by XLA,~as~reported in the tenth column of Table~\ref{table:benchmark}. For the other \todo{20} PPs, XLA does not have~any~performance improvement because of the small number of nodes in computation graphs. Thus, we consider these~\todo{20} PPs as not solved by XLA.  These results indicate that PPs in DL systems often cannot be eliminated by the compilation optimization techniques in XLA.

As shown in the last column of Table~\ref{table:benchmark}, \textsc{TensorFlow} documentation is only applicable~to \todo{12 (20.7\%)} PPs. We consider \textsc{TensorFlow} documentation as applicable as long as the documentation mentions the optimization solution of a PP. There are two main reasons that \textsc{TensorFlow} documentation is applicable to a small portion of PPs. The first is that performance characteristics, especially non-time characteristics, are hardly described~in~API~documentation.~The~second is that many PPs are caused by inefficient usages of multiple APIs, but API documentation is often focused on individual API usages. Although performance guide covers usages of multiple APIs, they only cover limited APIs such as \texttt{tf.data}. We consider these \todo{12} PPs as solved by \textsc{TensorFlow} documentation. These results show that \textsc{TensorFlow} documentation provides limited support for PPs.
 
\uline{\textbf{Summary.}} Efforts like profiling, compilation optimization~and~documentation have been devoted to optimizing~the~performance of DL systems from different perspectives. However, they provide limited capability in tackling PPs, potentially due to the lack~of~a~comprehensive understanding of PPs in DL systems.


\section{Implication, Application and Threat}

We discuss the implications for developers~and~researchers,~demonstrate one application to PP detection, and discuss the threats.

\subsection{Implications}

\textbf{Developers.} Our study reveals~the~common symptoms~of~PPs~that developers could pay attention~to~when testing and running their~DL systems for detecting potential PPs.~ Our study also identifies~the~common root causes of PPs that can~be~useful for developers~to~diagnose, debug or fix PPs. Our study also~captures the most bug-prone~or~bug-affecting stages where developers could~focus more efforts~on~to~provide the most benefit for PP introduction avoidance or performance optimization. Furthermore, our findings provide some development suggestions. Developers should carefully read the release note and API documentation of DL libraries to get familiar with the rich~set~of library APIs and their performance characteristics.~In this way, PPs caused by the most common root cause (i.e., API misuses) might~be reduced. Developers should also be systematically trained~to~have~a comprehensive understanding of computation graph to build efficient DL models. In this way, PPs caused by~the~second~most~common root cause (i.e., model construction) might be reduced.

\textbf{Researchers.} Our findings provide several implications on future research in three directions. First, \textit{intelligent techniques~for~high-performance DL system development} are needed. As developers~are~often unaware of library APIs that are specifically designed for high~performance or unaware of the performance characteristics~of~library~APIs, DL library API recommendation methods should be developed.~To~realize performance-aware API recommendation, a knowledge~graph of DL library APIs should be constructed based on release note,~API documentation and StackOverflow discussions with a specific focus on modeling performance characteristics of APIs and~performance differences across library versions. To locate and replace~inefficient code snippets written from scratch by developers, semantic~analysis techniques should be developed to determine their semantic~similarity to existing library APIs. Apart from such intelligent~techniques~at the code level, recommendation techniques~should be developed~to automatically suggest DL library versions, efficient~DL~models~and their parameters, and environment configurations.

Second, \textit{PP detection techniques} are needed. Half of the symptoms (i.e., \textit{Increasing Time Over Time},~\textit{Program Hang}, \textit{Out of Memory},~\textit{Memory Leak},~and \textit{Not Using GPU}) can be regarded as a credible~oracle~for detecting PPs in DL systems. Therefore, proactive~monitoring and prediction techniques should be developed to detect~PPs~as~early~as possible before these severe symptoms occur. DL systems exhibiting the~other symptoms are not guaranteed~to~contain~PPs~as~it~is~often not clear how much time or resources a DL system should consume to run without~a PP. To solve this performance~oracle~problem, one potential way is to design differential testing techniques~to~compare the performance of DL systems running with different DL~libraries, different DL models, or different hardware configurations.~However, it may incur too much overhead. Hence, another potential way~is~to design static techniques to model and estimate time cost~or~resource consumption of DL systems so that performance bottlenecks~can~be identified in advance before execution. During our manual analysis, we find that \textsc{TensorFlow} has some built-in mechanism~in~detecting PPs and recommending fixes~by~throwing a warning message, e.g., ``\textit{WARNING: tensorflow: multiprocessing can interact badly with TensorFlow, causing nondeterministic deadlocks. For high performance data pipelines tf.data~is~recommended}''. However, such warning~messages are only raised in \todo{3} of the PPs, indicating the preliminary support in PP detection due to symptom and root~cause diversity. Hence, built-in mechanisms in DL libraries should be further enhanced to detect PPs and recommend fixes. 


Third, \textit{PP localization techniques} are needed. Our study reveals~that the exposing stage of a PP is usually not the introducing stage.~For example, the location that throws the error message of an out~of~memory error is usually not the location of the root cause. Therefore,~it~is often challenging to localize PPs. During our manual analysis, we find that developers often use logs as the clue to locate PPs. Hence, automated log analysis techniques should be developed to smartly insert log statements into DL systems and locate potential~PPs~using log traces. Further, as API misuse is the most common root~cause of PPs, mining techniques should be designed to learn frequent API usage sequences and localize potential violations in DL systems.~API usage mining has been widely explored in traditional~systems~\cite{robillard2012automated}, but it is interesting to investigate how they are applicable to PPs in DL systems. From our experience, there are three challenges~to detect API-related PPs. First, due to the lack of effective type inference tool in Python, it is hard to precisely extract API usages from Python code. Second, as traditional API usage mining is not aware of performance characteristics of APIs, it is non-trivial~to~automatically determine the performance difference among mined API sequences. Third, it is difficult to detect PPs caused by \textit{Not Using Efficient APIs}, because the inefficient APIs that developers use are totally different from efficient APIs that should be used. Last but not the least, rule-based techniques should be developed to detect and localize PPs, considering the potentially large amount of PPs on StackOverflow or GitHub. The challenge is to automatically derive but not manually specify the rules.


\subsection{Application}

To demonstrate the usefulness of our findings, we implement~a~rule-based static checker, named \tool, to detect PPs in DL systems. \tool is implemented with two static analysis tools,~AST\footnote{https://docs.python.org/3/library/ast.html}~and Jedi\footnote{https://github.com/davidhalter/jedi/}. It currently supports three types of PPs whose detection~rules are manually derived from our empirical study (Sec.~\ref{sec:results}). 

\textbf{Checker 1: Repeated Node Creation.} Creating~the~same~nodes repeatedly to a computation graph is one of the common types~of~PPs under the root cause category of \textit{Confusion with Computation Graph}. \tool is designed to detect node creation APIs that are called~in loops with the same argument values; e.g., the two APIs \texttt{tf.matmul} and \texttt{optimizer.minimize} in Fig. ~\ref{fig:example-2}. Actually, it is similar to Loop Invariant Computation and Code Motion (LICM) optimization, which has been well studied in classic compilers~\cite{David1994LoopInvariant}. However, Grappler\footnote{https://tensorflow.org/guide/graph\_optimization}, the default graph optimizer in \textsc{TensorFlow} runtime, cannot eliminate this type of PPs although it has the loop optimizer. 
Notice that this type of PP has been reported in \cite{EmpiricalStudyTensorFlow2018}. However, to the best of knowledge, its detection has not been investigated in prior studies

To implement the checker, we first extract \textsc{TensorFlow} APIs~that may add computation graph nodes by parsing the \texttt{@tf\_export}~decorators in the source code of \textsc{TensorFlow} Python APIs\footnote{https://github.com/tensorflow/tensorflow/tree/r1.15/tensorflow/python/ops}. Then, we manually review these APIs to exclude APIs that actually~do~not~add nodes (e.g., \texttt{tf.assign}) or APIs that produce different values given~the same inputs (e.g., \texttt{tf.random.uniform}). Finally, we obtain~\todo{356}~APIs.

Our checker determines whether these~\todo{356}~APIs are called~with same argument values among loop~iterations.~To this end, it tracks variables that are changed among loop iterations, including the loop~control variable, variables that are assigned in the loop body but~are~defined outside the loop, and any variables that depend on them.~It~identifies APIs called without using changed variables as arguments~as~PPs. Our analysis is inter-procedural. If there are functions called~in~the loop, it passes changed variables to callee functions, analyzes changed variables in callee functions, and identifies APIs called without using changed variables as arguments in callee functions.


\textbf{Checker 2: Inefficient Order of \texttt{batch} and \texttt{map}.} As showed~in Fig.~\ref{fig:example-1}, calling \texttt{map} before \texttt{batch} is not efficient, and hence \texttt{batch} is suggested to be called before \texttt{map} to reduce the number of times the mapped function is called. To detect such API misuse of \texttt{batch}~and \texttt{map}, our checker first identifies \texttt{tf.Dataset} object, and then analyzes~the call sites to check whether \texttt{batch} is called after \texttt{map}. 

\textbf{Checker 3: Disabled Parallelism of \texttt{map} and \texttt{interleave}.} As listed in Fig.~\ref{fig:example-1}, calling \texttt{map} without setting its \texttt{num\_parallel\_calls} argument disables parallelism. It also holds for \texttt{interleave}.~To~detect such API misuse of \texttt{map} and \texttt{interleave}, our checker identifies \texttt{tf.Dataset} object, and analyzes the call sites to check~whether \texttt{map} and \texttt{interleave} are called without setting \texttt{num\_parallel\_calls}.

\begin{table}
\centering
\small
\caption{PP Detection Results of \tool}\label{table:detection}
\vspace{-10pt}
\begin{tabular}{|c|*{7}{C{2.2em}|}}
\hline
\multirow{2}{*}{\textbf{Checker}} & \multicolumn{3}{c|}{\textbf{Detected}} & \multicolumn{2}{c|}{\textbf{Confirmed}} & \multicolumn{2}{c|}{\textbf{Fixed}} \\\cline{2-8}
& \textbf{PP} & \textbf{Proj.}  & \textbf{FP} & \textbf{PP} & \textbf{Proj.} & \textbf{PP} & \textbf{Proj.} \\\hline\hline
Checker 1 & 77 & 49 & 15 & 20 & 14 & 7 & 4 \\\hline
Checker 2 & 195 & 68 & 0 & 52 & 18 & 0 & 0 \\\hline
Checker 3 & 216 & 66 & 0 & 33 & 18 & 20 & 10 \\\hline\hline
Total & 488 & 130 & 15 & 105 & 44 & 27 & 13 \\\hline
\end{tabular}
\end{table}

\textbf{Evaluation on Our Benchmark.} In our PP benchmark, four,~two and two PPs belong to the PP types targeted by the three checkers. Three, two and two of them were successfully detected by the three checkers. The only one false negative of \textbf{Checker 1} is caused~by~the incomplete type inference in Jedi. As reported~in~Sec.~\ref{sec:assessment},~\todo{\textsc{TensorFlow} Profiler is not applicable to these eight PPs. XLA is applicable to four, one and one of them, but fails to solve them. \textsc{TensorFlow} Documentation is applicable to zero, two and~two~of~them by only hinting the solution in API documentation or performance~guide.}

\textbf{Evaluation on GitHub Projects.} We used PyGitHub\footnote{https://github.com/PyGithub/PyGithub} to crawl \todo{1,108} GitHub repositories that used~\textsc{TensorFlow} and Python and had at least \todo{100} stars, and ran \tool~on these repositories.~We~reported detected PPs~as~issues~to~developers, and also manually~reviewed and verified all the detected PPs. As \textsc{TensorFlow} Profiler and XLA are dynamic analysis tools,~it~is difficult for us~to~properly configure and execute \todo{1,108} GitHub~repositories. \textsc{TensorFlow} Documentation only provides guidance but is not a tool.~Thus,~we~did not compare our checkers with~them~in this large-scale evaluation. The results are shown~in Table~\ref{table:detection}, where the statistics~about~detected, confirmed and fixed PPs are reported for each checker. 

Specifically, \textbf{Checker 1} detected \todo{77} PPs in \todo{49} projects. It detected \todo{15} false positives (i.e., the fourth column in Table~\ref{table:detection}). The reason is that we use lightweight heuristics to decide loop invariants~based on AST and Jedi, but do not use heavyweight data/control flow~analysis, for the scalability of our checker. \todo{20} PPs~in~\todo{14}~projects~have~been confirmed by developers, and \todo{7} of them in \todo{4} projects have been~fixed.
\textbf{Checker 2} detected \todo{195} PPs in \todo{68} projects with no false positive. \todo{52} PPs in \todo{18} projects have been confirmed by developers, but none~of them has been fixed. The reason is that   the fix requires extra effort~in vectorizing the mapped function (e.g., the \texttt{\_batch\_parser}~function in Fig.~\ref{fig:example-1}), which is non-trivial. In that sense,  automated~vectorization is required in \textsc{TensorFlow}, like auto-vectorization in LLVM\footnote{https://www.llvm.org/docs/Vectorizers.html\#slp-vectorizer}.
\textbf{Checker 3} detected \todo{216} PPs in \todo{66} projects with no false positive. \todo{33} PPs in \todo{18} projects have been confirmed by developers, while~\todo{20}~of them in \todo{10} projects have already been fixed. The projects~that~have confirmed/fixed our detected PPs~include popular ones like \textsc{Keras}, \textsc{TensorFlow Agents}, \textsc{TensorFlow Hub} and \textsc{Tensorforce}. Besides, we randomly sampled 5 PPs from the 7 and 20 fixed PPs for \textbf{Check 1} and \textbf{Checker 3} respectively, and measured the execution time of the buggy and fixed version. On average, the execution time was improved by 35.6\% and 20.4\% after fixing PPs, respectively.

\uline{\textbf{Summary.}} PP is~a~widespread problem~in~DL~systems, and rule-based PP detection is promising.  The three checkers in \tool detected \todo{488} PPs in \todo{130} projects with \todo{15} false positives. \todo{105} PPs in \todo{44} projects have~been~confirmed by developers, while~\todo{27}~of them in \todo{13} projects have~been~fixed by developers. 


\subsection{Threats}

We discuss the threats to our empirical study, PP benchmark,~and~detection approach. Our study investigates PPs in DL systems written with \textsc{TensorFlow} and \textsc{Keras}. Thus, it is not clear whether~our~findings can generalize to DL systems developed with other DL libraries like \textsc{PyTorch}. We believe it deserves a separate study to investigate differences across DL libraries. Further, our study~analyzes~PPs~from StackOverflow posts. However, GitHub is another~valuable~source~of PPs. It is interesting to further explore PPs from GitHub to strength our findings, which in fact requires large manual efforts as we spent \todo{six} person-months to analyze \todo{224} PPs. Our PP detection~results~on GitHub projects also indicate the potential applicability~of~our findings. Moreover, our study involves manual analysis on PPs, which may incur biases. To reduce them, two of the authors separately~analyzed PPs and a third author was involved to resolve disagreements.

Our benchmark consists of \todo{58} PPs, whose size,~to~be~honest,~is~not very large. However, considering the large human efforts involved~in constructing the benchmark, we believe it is acceptable. We are~still continuously enlarging our benchmark via reproducing those non-sampled PPs from the \todo{224} PPs and collecting PPs from GitHub.

Our rule-based static checker, \tool, currently only supports three types of PPs. Here, \tool is not designed to cover~all~type of PPs, but to demonstrate the potential of rule-based~PP~detection as well as the usefulness of our findings. We plan~to~manually~enrich the detection rules in \tool to support more PP types. In the long run, we hope to automatically learn~the~detection rules.

\section{Related Work}

We discuss the closely related work in understanding and analyzing deep learning bugs and performance problems.

\subsection{Deep Learning Bugs}\label{sec:dl-bug}

The recent success in applying deep learning techniques to a variety of domains has gained increasing interest in understanding~characteristics of bugs in deep learning systems. Zhang et al.~\cite{EmpiricalStudyTensorFlow2018} collected 175 bugs in deep learning systems developed~in~\textsc{TensorFlow}~from StackOverflow posts and GitHub commits. They~analyzed the symptoms and root causes of these bugs, and explored~the~challenges~and strategies in bug detection and localization. Islam~et~al.~\cite{Islam2019}~and~Humbatova et al.~\cite{humbatova2019taxonomy} expanded the scope of Zhang et al.'s study~to~include more deep learning libraries. Islam et al.~\cite{Islam2019} analyzed~types, root causes, impacts and pipeline stages of 970 bugs in deep learning systems written in \textsc{Caffe}, \textsc{Keras}, \textsc{TensorFlow}, \textsc{Theano}~and~\textsc{Torch}, while Humbatova et al.~\cite{humbatova2019taxonomy} constructed a taxonomy of bugs in deep learning systems that use \textsc{TensorFlow}, \textsc{Keras} and \textsc{PyTorch} based on manual analysis of 375 bugs and interviews with 20 developers. In their follow-up work, Islam et al.~\cite{islam2020repairing} analyzed bug fix patterns. Kim et al.~\cite{kim2021denchmark} built a benchmark of 4,577~bugs from~193~deep learning systems. Differently, Jia et al.~\cite{jia2020empirical} explored the symptoms, root causes and locations of 202 bugs in the \textsc{TensorFlow} library.

Apart from the studies that are focused on a general scope~of~bugs in deep learning systems, several recent studies have targeted more specific bugs. Zhang et al.~\cite{zhang2020empirical} studied failures of deep learning~jobs that are running on a remote, shared platform in Microsoft. Chen~et al.~\cite{chen2021empirical} investigated faults related to the deployment of deep learning models to mobile devices. Zhang et al.~\cite{zhang2021autotrainer} summarized five common training problems in deep learning systems, and developed~a tool to automatically detect and repair training problems. Wan et al.~\cite{MLAPI2021} studied API misuses when deep learning systems use cloud AI services, summarized eight misuse patterns, and developed static checkers to automatically detect some of the misuse~patterns. Huang et al.~\cite{dependency_bug} explored dependency bugs across the DL stack.

Some of these studies reveal some partial characteristics of performance problems in deep learning systems. For example, Zhang et al.~\cite{EmpiricalStudyTensorFlow2018} and Islam et al.~\cite{Islam2019} respectively recognized low efficiency and hang~as a symptom of deep learning bugs.~Zhang~et~al.~\cite{zhang2020empirical}~identified GPU out of memory as a failure category of deep learning~jobs. Chen et al.~\cite{chen2021empirical} recognized memory and speed issues~as~two types~of faults in the inference stage of deployment process.~Wan~et~al.~\cite{MLAPI2021}~derived four performance-related API misuse patterns of cloud~AI~services. \todo{Despite these efforts, there still lacks a comprehensive study~to understand characteristics of performance problems in deep~learning systems, and thus our study aims to bridge this knowledge~gap and raise the awareness of performance problems in DL systems.}

Besides, some studies have explored general problems~and~challenges in developing and deploying deep learning systems.~For~example, Guo et al. \cite{guo2019empirical} measured the accuracy and performance differences across four deep learning libraries. Zhang~et~al.~\cite{zhang2019empirical}~identified seven kinds of frequently asked deep learning questions~in~StackOverflow, and analyzed their resolution difficulty and root causes. Han et al.~\cite{han2020programmers} explored the topics that developers discuss when~developing deep learning systems. Chen et al.~\cite{chen2020comprehensive} built a taxonomy~of challenges in deploying deep learning systems to different platforms through manual analysis of StackOverflow posts. Pham~et~al.~\cite{pham2020problems} measured accuracy variance in training deep learning systems.~Cummaudo et al.~\cite{Cummaudo2020} studied pain-points that developers face~when~using cloud services of computer vision by mining StackOverflow~posts. 

Although these studies are not designed for deep learning~bugs,~they shed light on debugging and bug detection in deep learning~systems. Specifically, Guo et al. \cite{guo2019empirical} reported performance differences~in~terms of time cost and memory consumption when trained~deep~learning models are migrated or quantized to different mobile devices~and~web browsers, and called for performance optimization and testing techniques. Zhang et al.~\cite{zhang2019empirical} summarized performance as a category of frequently asked deep learning questions in StackOverflow,~and recognized that performance questions are the most difficult~to~answer. \todo{Our study is inspired by these studies to systematically characterize performance problems in deep learning systems.}

Moreover, some advances have been made to detect deep learning bugs. For example, Zhang et al.~\cite{zhang2020detecting} developed a static analysis~approach to detect numerical bugs in neural architectures based on abstract interpretation. Lagouvardos et al.~\cite{lagouvardos2020static} proposed~a~static~analysis to detect shape incompatibility errors in \textsc{TensorFlow} programs, while Verma and Su~\cite{Verma2020} proposed a dynamic abstract interpreter~to catch such errors. Wardat et al.~\cite{wardat2021deeplocalize} developed~a~dynamic~analysis approach to locate faults in deep neural networks.~In~addition,~great efforts have been devoted to testing deep learning systems (e.g.,~\cite{pei2017deepxplore, tian2018deeptest, ma2018deepgauge, odena2019tensorfuzz, xie2019deephunter, sun2018concolic, kim2019guiding}) and deep learning libraries~(e.g.,~\cite{pham2019cradle, nejadgholi2019study, guo2020audee, wang2020deep, wang2021automatic, zhang2021predoo}) for quality assurance. Zhang et al.~\cite{zhang2020machine} presented~a~comprehensive survey of work in this direction. \todo{However, little attention has been received to detecting and testing performance problems in deep learning systems, and our study sheds light on this area.}

\subsection{Performance Problems} 

Many empirical studies have characterized performance~problems~from different perspectives (e.g., root causes, discovery, diagnosis, fixing and reporting) for desktop or server applications~\cite{UnderstandingDetectingRealworld2012, zaman2012qualitative, nistor2013discovering, Song2014, HowArePerformance2020}, highly configurable systems~\cite{EmpiricalStudyPerformance2016, he2020cp}, mobile applications~\cite{liu2014characterizing, linares2015developers}, database-backed web applications~\cite{yang2018not, yang2019view}, and~JavaScript~systems \cite{selakovic2016performance}. They shed light on potential directions on performance 
analysis (e.g., detection, profiling and testing).~\todo{Our study~is~the first~to understand performance problems in deep learning systems,~which~differs from traditional systems on the programming paradigm.}

Advances (e.g., \cite{ammons2004finding, cito2019interactive, han2012performance, curtsinger2015coz}) have been made to identify~general performance problems with dynamic profiles from production~runs. A large body of work has designed pattern-based~methods~to~detect specific performance problems, e.g., reusable/cacheable data (e.g., \cite{bhattacharya2011reuse, Toffola2015, nguyen2013cachetor}), inefficient/redundant loops (e.g., \cite{dhok2016directed, nistor2013toddler, song2017performance, nistor2015caramel}),~and~inefficient collections (e.g., \cite{Jung2011, Shacham2009, Xu2010}). Besides, a  lot of techniques~have been proposed for performance testing, i.e., generating test cases~to trigger worst-case performance (e.g., \cite{burnim2009wise, luckow2017symbolic, lemieux2018perffuzz, petsios2017slowfuzz, wei2018singularity}) and find~performance problems (e.g., \cite{grechanik2012automatically, shen2015automating, tizpaz2020detecting}). Another line of work~is performance profiling technique to identify hot paths (e.g., \cite{ball1996efficient, Duesterwald2000, Larus1999}) and fit a performance model to the input size (e.g., \cite{Coppa2012, goldsmith2007measuring, Zaparanuks2012}).~\todo{These performance analysis approaches are designed for traditional systems, and cannot be directly applied to deep learning systems.}

Recently, some performance analysis approaches have been proposed for deep learning systems. For example, Qi et al.~\cite{qi2016paleo} modeled and estimated time cost of training deep neural networks,~while~Gao et al.~\cite{gao2020estimating} estimated GPU memory consumption. \todo{Such estimation techniques are useful to find potential performance problems~in~advance.}
Liu et al.~\cite{liu2019performance} measured the performance~of~training~deep~learning models on mobile devices, while Ma et al.~\cite{ma2019moving}~compared time cost of JavaScript-based deep learning libraries when running deep learning tasks~in~browsers. \todo{These studies empirically demonstrate the performance differences.} To reduce memory usage of deep~neural networks,~Rhu et al.~\cite{rhu2016vdnn} developed a dynamic memory manager to~virtualize memory~usage, while Wang et al.~\cite{wang2018superneurons} proposed~a~dynamic GPU memory~scheduler. To make deep learning models~efficient, Han et al.~\cite{Han2016} used pruning and quantization to compress models, Yan~et~al.~\cite{yan2015performance} used~a~performance model~to estimate the time of distributed model training~and find the optimal distributed configuration, and Menghani~\cite{Menghani2021} presented a survey in this area. \todo{These approaches are system-level performance optimization techniques, while \tool is at the source code level.}~\todo{Despite these efforts, the characteristics of performance problems in deep learning systems are still unclear, and our study fills this gap.}
 

\section{Conclusions}

We present the first comprehensive~study~to~characterize PPs~in~DL systems written~in~\textsc{TensorFLow}~and \textsc{Keras}, and build the first~benchmark of PPs in DL systems to assess existing~approaches in tackling them. Further,~we develop a static checker \tool to detect~three types of PPs,~and detect many new PPs in GitHub projects.



\section{Data-Availablity Statement}
All the study data and source code of \tool are available at \cite{zenodo} to foster future research.


\begin{acks}
This work was supported by the National Key R\&D Program of China (2021ZD0112903).
\end{acks}

\balance

\bibliographystyle{ACM-Reference-Format}
\bibliography{src/reference}

\end{document}